\documentclass[pre,twocolumn,floatfix]{revtex4-1}
\usepackage[pdftex]{color,hyperref}

\usepackage{amsmath,amsfonts,amssymb}
\usepackage{subcaption}
\usepackage{xcolor}

\usepackage{graphicx}

\begin{document}

\title{Multichannel Social Signatures and Persistent Features of Ego Networks}
\author{S. Heydari}
\affiliation{Department of Computer Science, Aalto University, P.O. Box 14400, FI-00076 AALTO, Finland}
\email{sara.heydari@aalto.fi}
\author{S.G.B. Roberts}
\affiliation{School of Natural Sciences and Psychology, Liverpool John Moores University, Byrom Street, Liverpool L3 3AF, UK}
\author{R.I.M. Dunbar}
\affiliation{Department of Experimental Psychology, University of Oxford, Oxford OX1 3UD, UK}
\author{J. Saram\"aki}
\affiliation{Department of Computer Science, Aalto University, P.O. Box 14400, FI-00076 AALTO, FInland}
\affiliation{Helsinki Institute of Information Technology HIIT, Aalto University, Finland}
\date{\today}

\begin{abstract}

The structure of egocentric networks reflects the way people balance their need for strong, emotionally intense relationships and a diversity of weaker ties. Egocentric network structure can be quantified with 'social signatures', which describe how people distribute their communication effort across the members (alters) of their personal networks.
Social signatures based on call data have indicated that people mostly communicate with a few close alters; they also have persistent, distinct signatures. To examine if these results hold for other channels of communication, here we compare social signatures built from call and text message data, and develop a way of constructing mixed social signatures using both channels. We observe that all types of signatures display persistent individual differences that remain stable despite the turnover in individual alters. We also show that 
call, text, and mixed signatures resemble one another both at the population level and at the level of individuals. The consistency of social signatures across individuals for different channels of communication is surprising because the choice of channel appears to be alter-specific with no clear overall pattern, and ego networks constructed from calls and texts overlap only partially in terms of alters. These results demonstrate individuals vary in how they allocate their communication effort across their personal networks and this variation is persistent over time and across different channels of communication.

\end{abstract}

\maketitle

\section{Introduction}

Social relationships that are strong and supportive are fundamentally important for health and well-being, in both humans and other primates \cite{house1988,lyubomirsky2005,wittig2008,holtlunstad2010,manninen2017}. While close, emotionally intense relationships provide support and cohesion, weaker ties have been associated with the benefits of diversity and access to resources outside one's everyday social circles~\cite{Granovetter,Burt}. At the same time, maintaining social ties comes at a cost: time and cognitive resources are finite \cite{miritello2013time,miritello2013limited}. This cost is particularly high for close relationships~\cite{Roberts2009}. As a result, personal networks typically have only a few close ties and many weak ties. This is visible both at the level of entire social networks \cite{Onnela2007} as well as in how people structure their personal networks\cite{saramaki2014persistence}.

The way people balance their need for strong, cohesive ties and weak ties that lead outside their closest network is reflected in so-called \emph{social signatures}. The social signature of an ego measures the fraction of communication targeted at alters of each rank, when the alters are ranked according to this fraction. Social signatures therefore quantify rank-frequency relationships of alters in egocentric networks.
In Ref.~\cite{saramaki2014persistence}, it was shown that people place their mobile telephone calls to their alters very unevenly across their ego networks, so that a few closest alters get a disproportionate fraction of calls. This is reflected in social signatures that typically decay slower than exponentially. It was also shown in Ref.~\cite{saramaki2014persistence} that 
each individual has their own, distinctive social signature that  persists in time, even when there is a large amount of turnover in the ego network. Similar observations were made in Ref.~\cite{centegheller2017} with a different dataset on mobile telephone calls.

However, social relationships are shaped and maintained through a diversity of communication channels~\cite{Vlahovic2012,reid2005,nanavati2008,wang2013,zignani2014,quadri2014}. People do not use these channels uniformly -- rather, the choice of channel depends on many factors. These include the type of relationship (nature of social tie), general channel preferences, the time of the event (social norms) and the reason for communicating; see, e.g.,~\cite{reid2005} on why texters text. To examine if the properties of social signatures are generalizable and genuine features of egocentric networks, it is therefore important to look at data from multiple channels of communication, both separately and together.
Combining information on different channels can, however, be problematic because of their intrinsic differences. For example, the number of calls or their total duration is typically used as a proxy for tie strength in mobile telephone call data~\cite{saramaki2015seconds}. But text messages, another common form of communication via mobile devices, have no duration, and the number of text messages between an ego-alter pair is not directly comparable to the number of calls between that pair. While one call can be thought to represent one conversation, one text message is typically only a part of a longer conversation.  

In this paper, we study social signatures that are based on calls, texts, and both. To this end,
we develop a way of constructing weighted ego networks from time-stamped communication data that makes different channels more comparable (see Fig.~1), and also allows for the construction of mixed social signatures based on both call and text message data. We apply this method to two data sets on mobile telephone communication, and observe that both single-channel and mixed signatures are persistent over time, as observed earlier for calls-only signatures. We also observe that the call and text signatures are surprisingly similar for each ego. This is unexpected, because at the same time, the call and text networks of most egos overlap only partially, and there are no clear patterns of channel preference: the choice of channel appears independent of alter rank in mixed signatures.

\section{Materials and Methods}

\subsection{Datasets}

\begin{table*}[t!]
\caption{The two data sets used in this study. NCPM = number of calls per user per month; NTPM = number of text messages per user per month.}
\centering
\begin{tabular}{ccc}
\hline
    & DS 1 & DS 2 \\ 
\hline
Number of active users  & 506330  & 24  \\
Length of data-collection period    & 7 months & 18 months  \\
10-percentile of NCPM & 43 & 67 \\
Median of NCPM & 83 & 127  \\
90-percentile of NCPM & 194 & 278  \\
10-percentile of NTPM & 20 & 105 \\
Median of NTPM & 45 &  317 \\
90-percentile of NTPM & 131 & 2019 \\
\hline
\end{tabular}
  \label{table:datasets}
\end{table*}

We use two data sets of mobile telephone calls and text messages (see Table~\ref{table:datasets}). \emph{Data set DS1} comprises the Call Detail Records (CDRs) for calls and text messages of the anonymized customers of a mobile operator in an European country, collected over 7 months (see, e.g., \cite{Karsai2011,kivela2012multiscale}). We applied an activity threshold and retained only users with more than $20$ calls and more than $7$ text messages per month, retaining 506,330 users. \emph{Data set DS2} contains the times and recipients of outgoing calls and text messages for 24 students in the UK~\cite{Roberts11,saramaki2014persistence}. The data collection period is 18 months, during which the students graduated from high school and moved on to University or work.

As our aim was to construct social signatures and study their persistence in time, we divided both data sets into two equal-sized consecutive time intervals (3.5 months each for DS1 and 9 months for DS2); this was for being able to compare the stability of the shapes of the signatures for the first and second halves. The choice to split into two was merely for convenience; please note that in ~\cite{saramaki2014persistence}, DS2 was analyzed using three intervals, yielding similar results for calls.

\subsection{Constructing egocentric networks and social signatures}

Social signatures are calculated from weighted egocentric networks, where the link weights represent the amount of communication between the focal ego and the ego's alters. Social signatures measure the fraction of communication to alters of each rank, when the alters are ranked according to this fraction. In Ref.~\cite{saramaki2014persistence}, the number of outgoing calls that took place during the data collection period were used as weights. However, when there are multiple channels, the question of how to define weights is not straightforward. The simplest solution would be to use the number of communication events as the weight for all channels. However, this is  problematic. In our case of calls and texts, as disussed above the numbers of calls and texts cannot be directy compared. One call can be associated with one conversation, while one conversation by texting may amount to a large number of individual text messages. 

\begin{figure*}[ht!]
\includegraphics[width=0.65\linewidth]{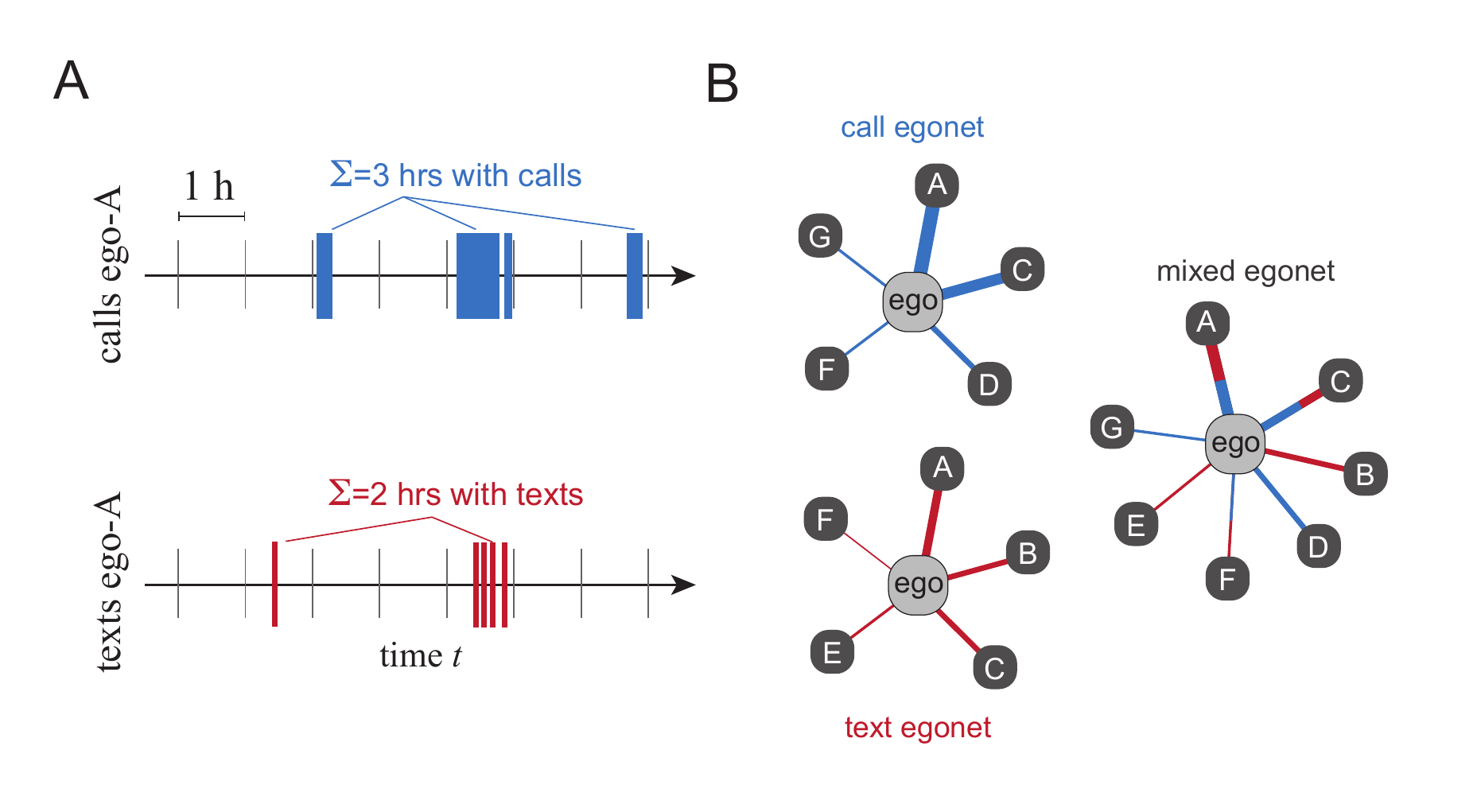}
\caption{Constructing egocentric networks from calls and texts using time-binned weights. A) The timelines corresponding to each of the ego's alters are divided into bins--we use bins that span one hour. Then, the number of bins with at least one communication event is computed. These numbers are used as link weights for egocentric networks (panel B). For the mixed networks, the link weights represent the number of bins where either calls, texts, or both are taking place.
}\label{schematic}
\end{figure*}

Here, our aim is to make the channels more comparable by focusing on their timelines and coarse-graining events in time. We do this as follows: we take the timeline of each ego-alter link, and divide it into time bins of one hour. Note that one hour has been chosen for convenience and to be clearly longer than the time scale of tens of seconds to minutes associated with correlated calls or texts~\cite{Karsai2011,backlund2014}).
Then, for both calls and texts, we count the number of bins which contain at least one communication event (see Fig.~\ref{schematic}A). Thus we count the number of one-hour time bins in which at least one communication activity takes place. These counts are then used as link weights for the egocentric networks: \emph{e.g.}, a weight of $w=5$ indicates that there were five hours where there was call activity with the alter. Calls that begin in one time bin and stretch along several time bins contribute accordingly to several units of weight. Defining link weights on the basis of time bins also makes it possible to construct \emph{mixed} link weights, as one can count the number of time slots where either at least one call OR one text message took place. 
An advantage of this method is that it can be used to calculate link weights that quantify the amount of communication or social interaction in any channel, as long as the time stamps of interaction events are available for each ego-alter link.  

With the time-bin-based weights, social signatures are calculated as in \cite{saramaki2014persistence}: for each egocentric network, alters are ranked according to their link weight, and the fraction of link weight out of the sum of all link weights is computed as function of alter rank. The social signature of ego $i$ then reads
\begin{equation}
\sigma_i=\{(w_{i1}/\sum_{j=1}^{k_i}{w_{ij}}),\ldots,(w_{ik_i}/\sum_{j=1}^{k_i}{w_{ij}})\},\label{eq:signature}
\end{equation}
where the alters $j$ are sorted by weight in decreasing order and $k_i$ is the degree (number of alters) of $i$.

\subsection{Comparing Social Signatures}
In order to determine the persistence of social signatures in time, a way of comparing their shapes and measuring the similarity or difference between two given signatures is needed. In (\cite{saramaki2014persistence}), the Jensen-Shannon divergence \cite{lin1991divergence} (JSD) was used for comparing pairs of social signatures and we also use the JSD in this analyses. The JSD is defined as:
\begin{equation}
JSD_{(\sigma_1, \sigma_2)} = H(\frac{1}{2}\sigma_1 + \frac{1}{2}\sigma_2) - \frac{1}{2}[H(\sigma_1)+H(\sigma_2)],
\end{equation}
where $\sigma_1$ and $\sigma_2$ are two social signatures, as defined in Eq.~\ref{eq:signature}, and $H(\sigma)$ is the Shannon entropy of $\sigma$.

The Jensen-Shannon divergence is a generalized form of the Kullback-Leibler divergence. The square root of the Jensen-Shannon divergence can be used as a distance function. Because the JSD can deal with zero probabilities, it allows us to compare social signatures of different lengths, that is, signatures computed from egocentric networks with different numbers of alters, $k_1 \neq k_2$. To compare two signatures of lengths $k_1$ and $k_2$ where $k_2 > k_1$, we append zero entries ($w_{ij}=0$) to the shorter signature for $k_1 <j \leq k_2$ so that both signatures are of equal length.

The overlap of two sets of alters in a pair of egocentric networks can be measured by the Jaccard coefficient
\begin{equation}
J(\sigma_1,\sigma_2) = \frac{|\sigma_1 \cap \sigma_2|}{|\sigma_1 \cup \sigma_2|},
\end{equation}
where $\sigma_1$ and $\sigma_2$ are the social signatures corresponding to the networks. As an example, the Jaccard coefficient between the call signature and text signature of an ego is defined as the number of alters the ego has contacted by both call and text in the period divided by number of alters the ego has contacted by call or text. If there is complete overlap between the alters contacted by call and text, then J = 1. If there is no overlap between alters contacted by call and text, then J = 0.

\section{Results}
\subsection{Single-Channel and Mixed Signatures Are Persistent}

We begin our analysis by demonstrating that all three types of signatures -- call, text, and mixed -- are persistent at the level of individuals, as was shown for call-based signatures for DS2 in Ref.~\cite{saramaki2014persistence}. Here, we define persistence as the social signature retaining its shape over time, with individual level variation in JSD that is smaller than the average JSD between signatures in the whole population.

To examine this persistence, we divide the data collection periods of the two sets into two intervals of 3.5 and 9 months for sets DS1 and DS2, respectively. We then calculate the weighted egocentric networks for each ego in each interval. As explained in detail above, we use the number of one-hour time bins with calls, texts, or either for determining the link weights between the ego and alters. We compute the social signatures for each egocentric network and each interval by ranking alters according to their weight and calculating the fraction of weight at each rank. Following  Ref.~\cite{saramaki2014persistence}, we then calculate self-distances by computing the JSD between an ego's own signatures in consecutive intervals. We also calculate reference distances by computing the average JSD between the signature of the ego and those of all other egos. We repeat these calculations for both channels (calls and texts) as well as mixed networks (calls and texts).

\begin{figure*}
\centering
\begin{subfigure}{.45\textwidth}
  \centering
  \includegraphics[width=\linewidth]{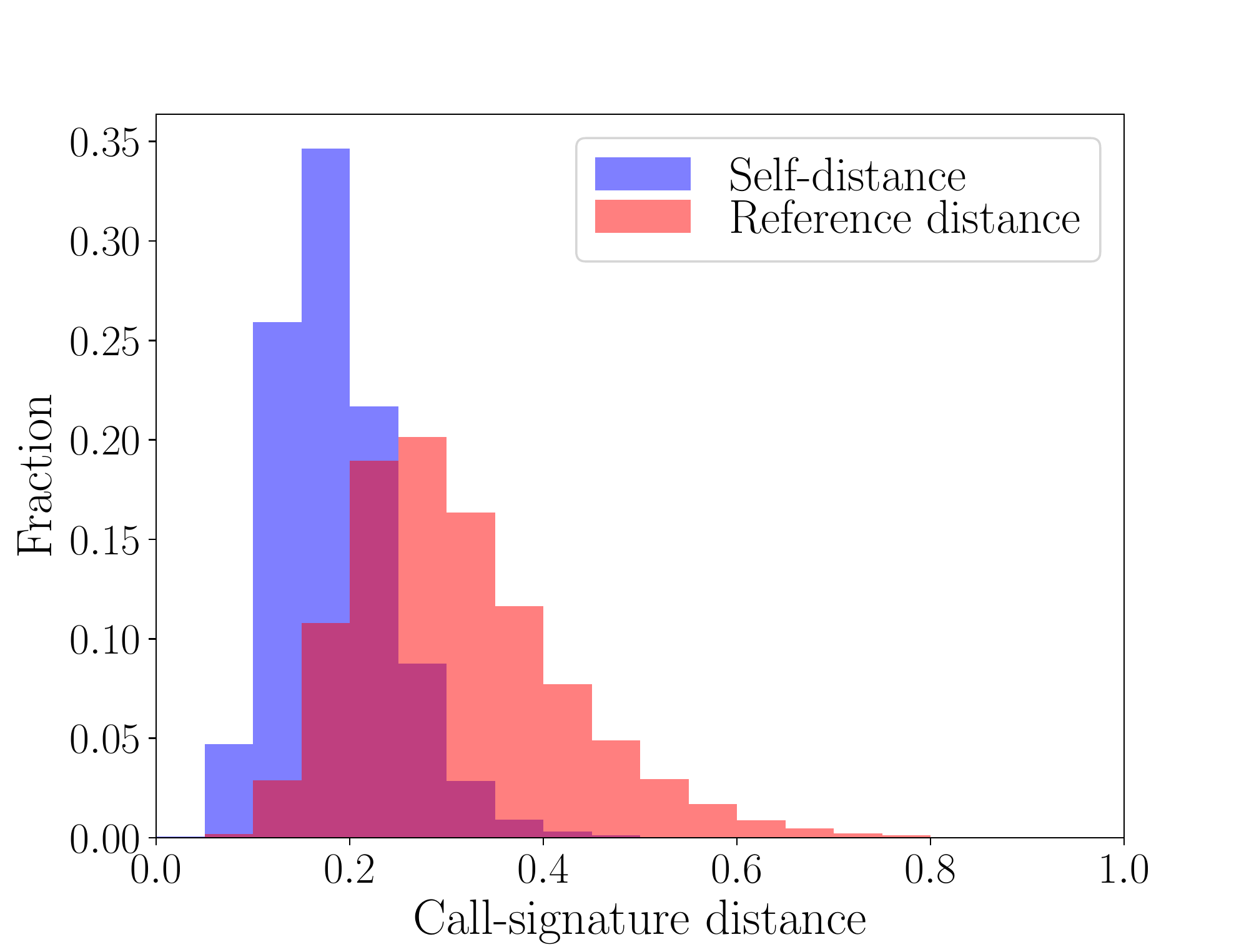}
  \caption{}
  \label{fig:big_call_persistence}
\end{subfigure}%
\begin{subfigure}{.45\textwidth}
  \centering
  \includegraphics[width=\linewidth]{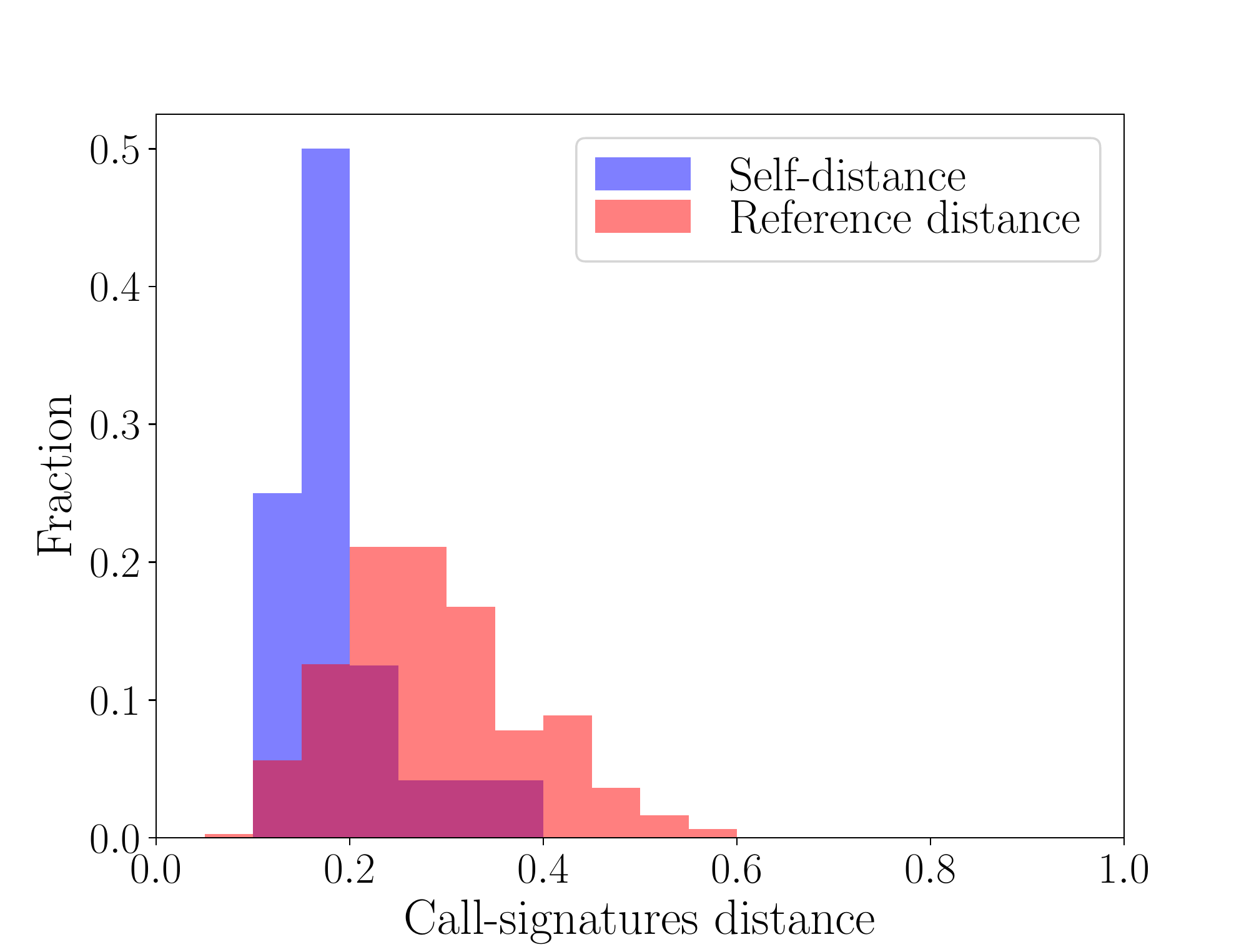}
  \caption{}
  \label{fig:oxford_call_persistence}
\end{subfigure}
\begin{subfigure}{.45\textwidth}
  \centering
  \includegraphics[width=\linewidth]{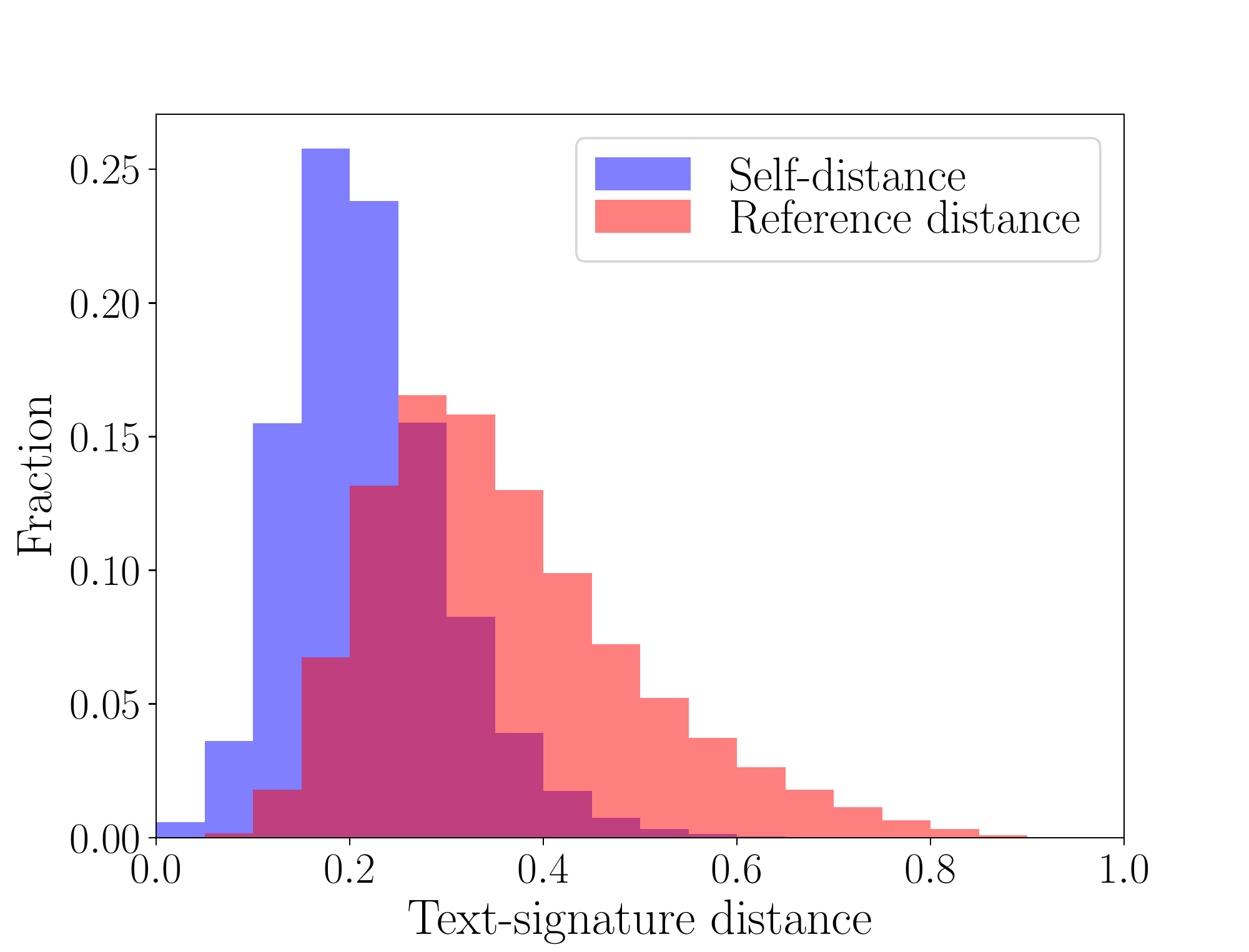}
  \caption{}
  \label{fig:big_text_persistence}
\end{subfigure}%
\begin{subfigure}{.45\textwidth}
  \centering
  \includegraphics[width=\linewidth]{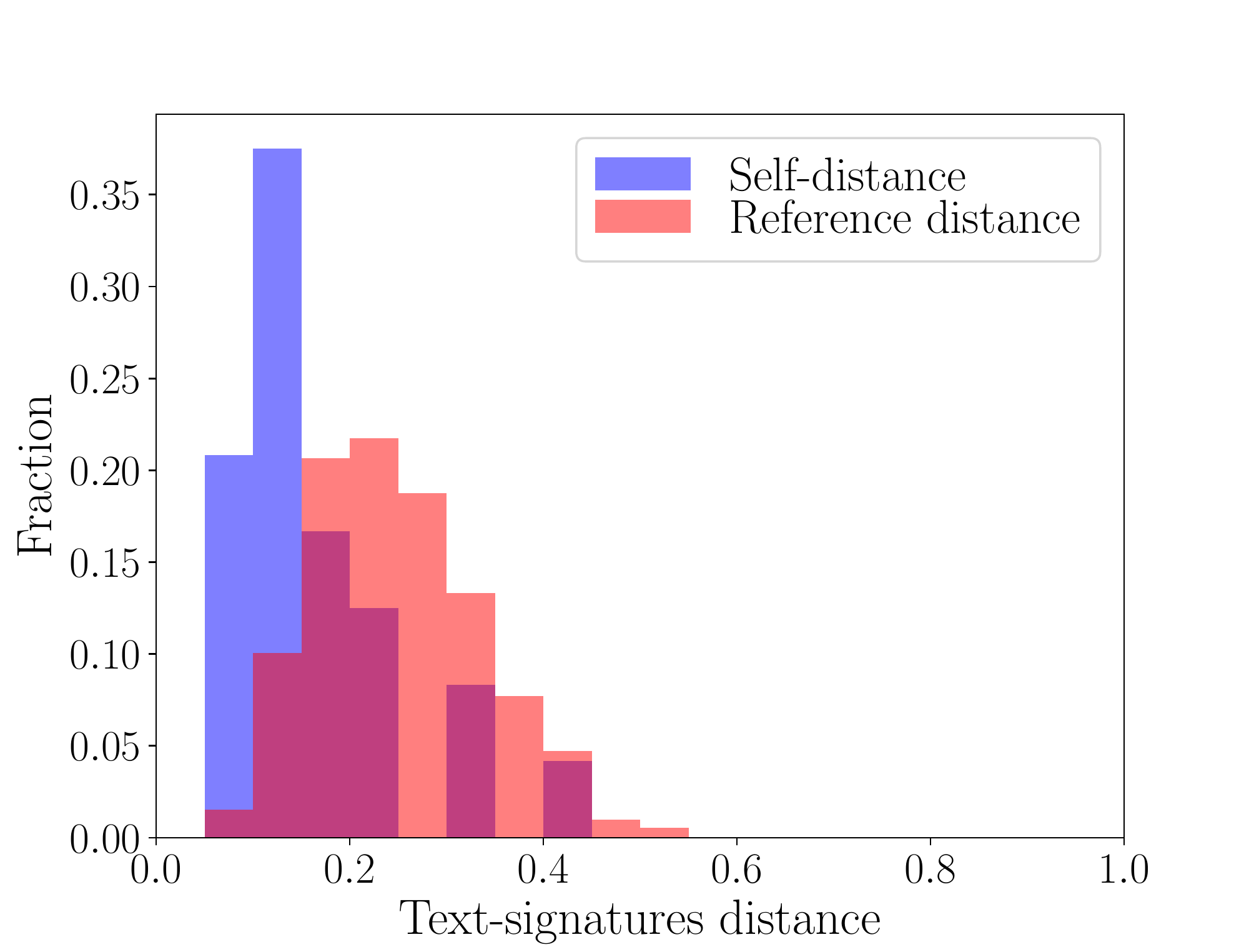}
  \caption{}
  \label{fig:oxford_text_persistence}
\end{subfigure}
\begin{subfigure}{.45\textwidth}
  \centering
  \includegraphics[width=\linewidth]{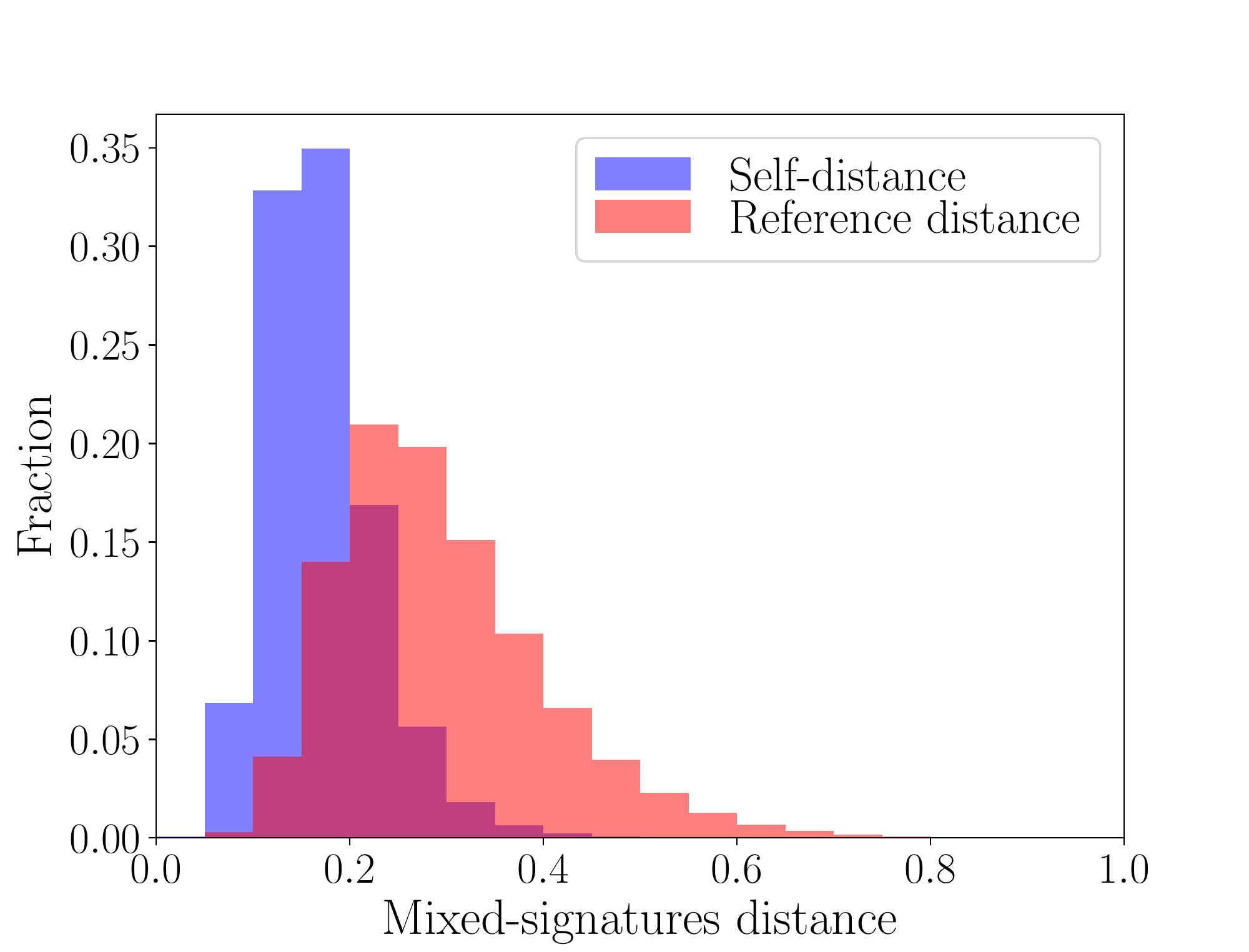}
  \caption{}
  \label{fig:big_mixed_persistence}
\end{subfigure}%
\begin{subfigure}{.45\textwidth}
  \centering
  \includegraphics[width=\linewidth]{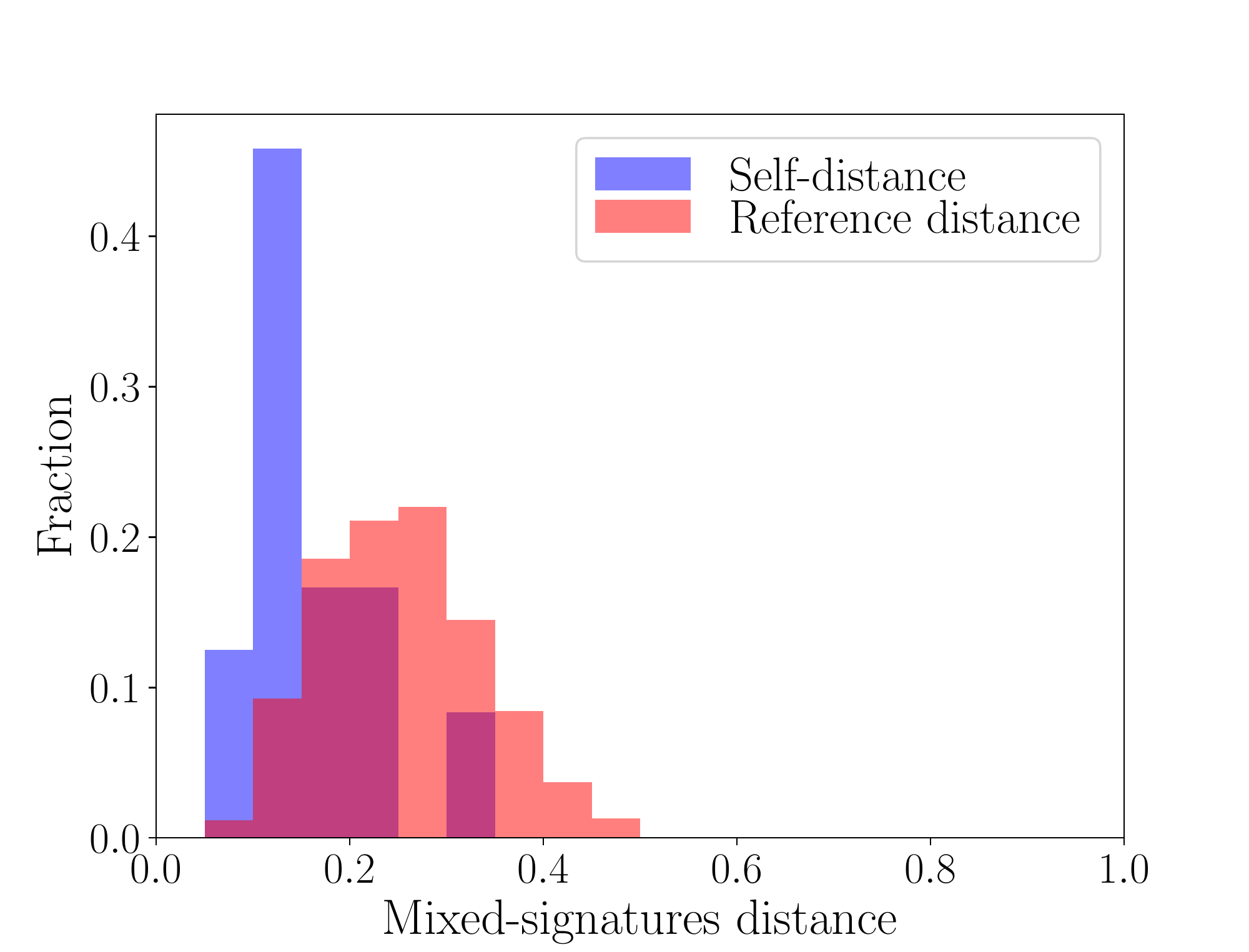}
  \caption{}
  \label{fig:oxford_mixed_persistence}
\end{subfigure}
\caption{\label{fig:persistence_call_sig}
Social signatures are persistent at the individual level. This holds for both channels (calls and texts) as well as mixed signatures combining both. Panels (a), (c), (e): Dataset 1 (the large dataset), Panels (b), (d), (f): Dataset 2 (students). The distributions of distances between social signatures of each ego in two consecutive equal-sized intervals are shown in blue (self--distances). The reference distributions of distances between signatures of different egos are shown in red (reference distances). Comparing the distributions of self--distances with reference distances verifies the persistence of call, text and mixed signatures, as self-distances are on average smaller than reference distances.}
\end{figure*}

The distributions of self and reference distances of call, text, and mixed signatures are displayed in Fig.~\ref{fig:persistence_call_sig}. For all three types of signatures and for both data sets, the bulk of the distribution of self-distances clearly lies below the reference distances. Self-distances are on average smaller than the distance between the signatures of two random egos, and there is less spread in their distribution. These differences in the distributions of Fig.~\ref{fig:persistence_call_sig} indicate that the changes of an ego's signature in time are smaller than the variation of signature distances in the population, whether calculated from calls or texts or both. This means that the individual differences in signature shapes are a real feature of the egocentric networks instead of random variation resulting in noisy, unstable signatures. 
The persistence of social signatures is therefore not only a feature of egocentric networks built from phone call data, but a more general phenomenon. 

\subsection{Single-Channel and Mixed Signatures Have Similar Shapes, Even at the Ego Level}

\begin{figure*}
\centering
\begin{subfigure}{.45\textwidth}
  \centering
  \includegraphics[width=\linewidth]{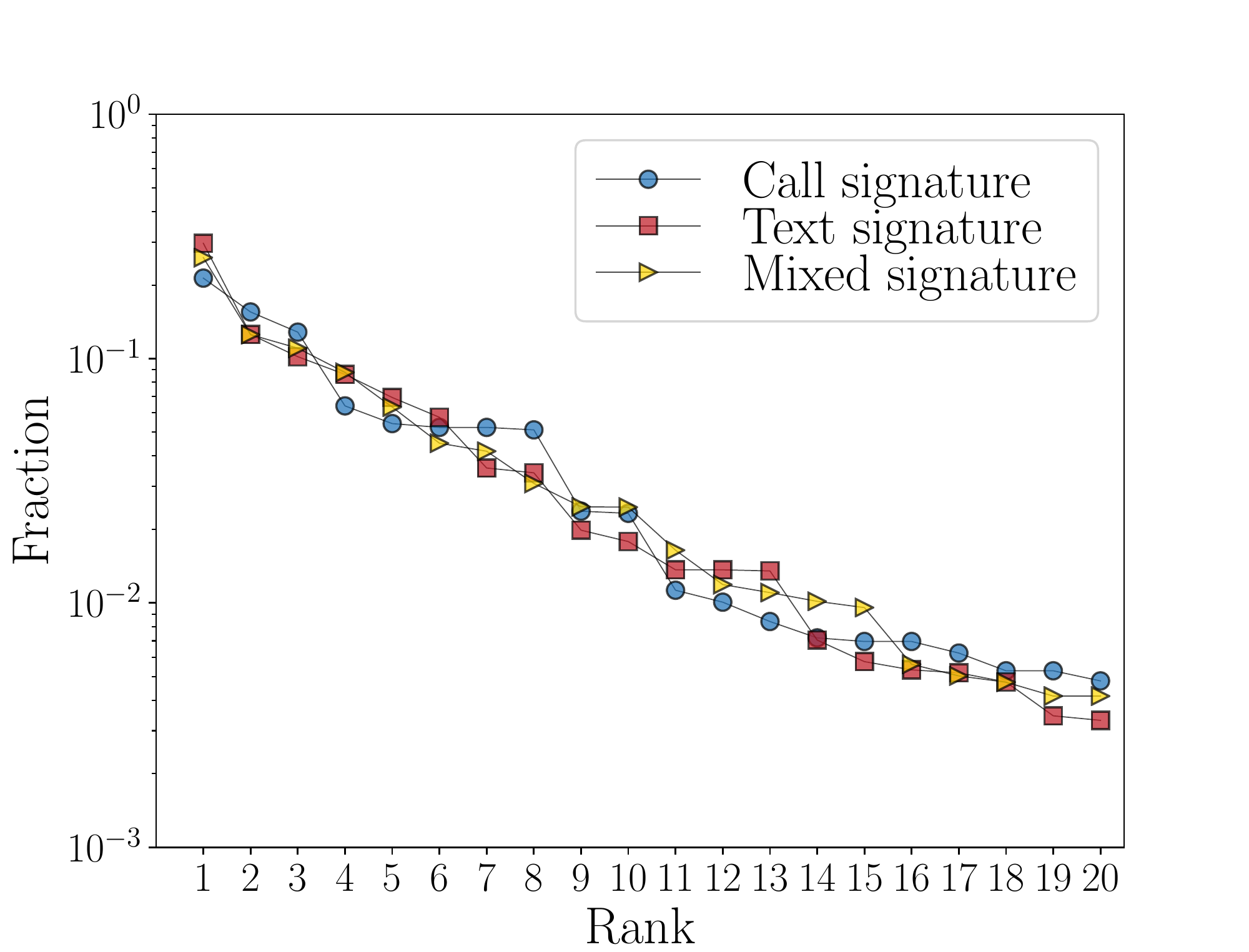}
  \caption{}
  \label{fig:example_sig}
\end{subfigure}%
\begin{subfigure}{.45\textwidth}
  \centering
  \includegraphics[width=\linewidth]{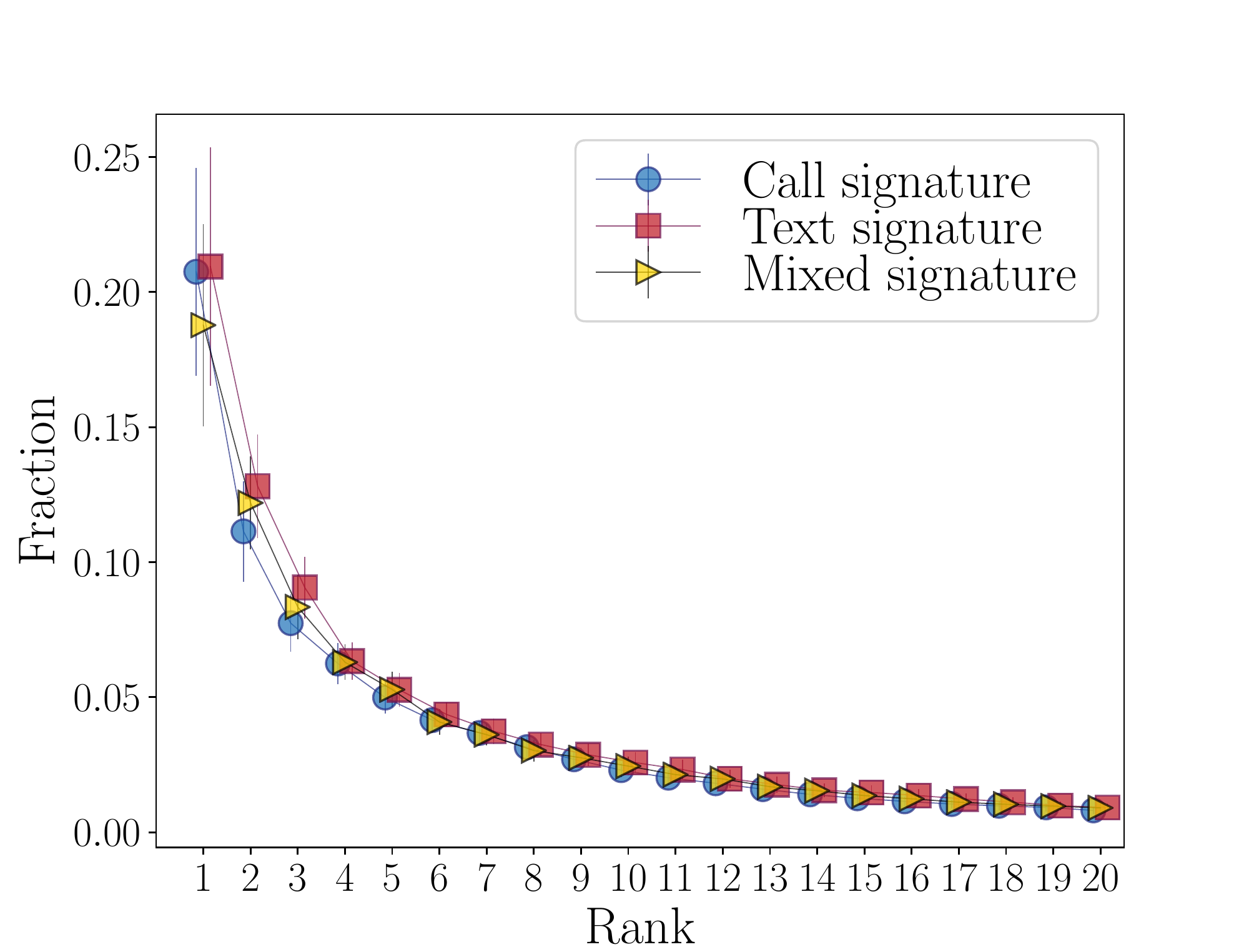}
  \caption{}
  \label{fig:average_sigs}
\end{subfigure}
\begin{subfigure}{.45\textwidth}
  \centering
  \includegraphics[width=\linewidth]{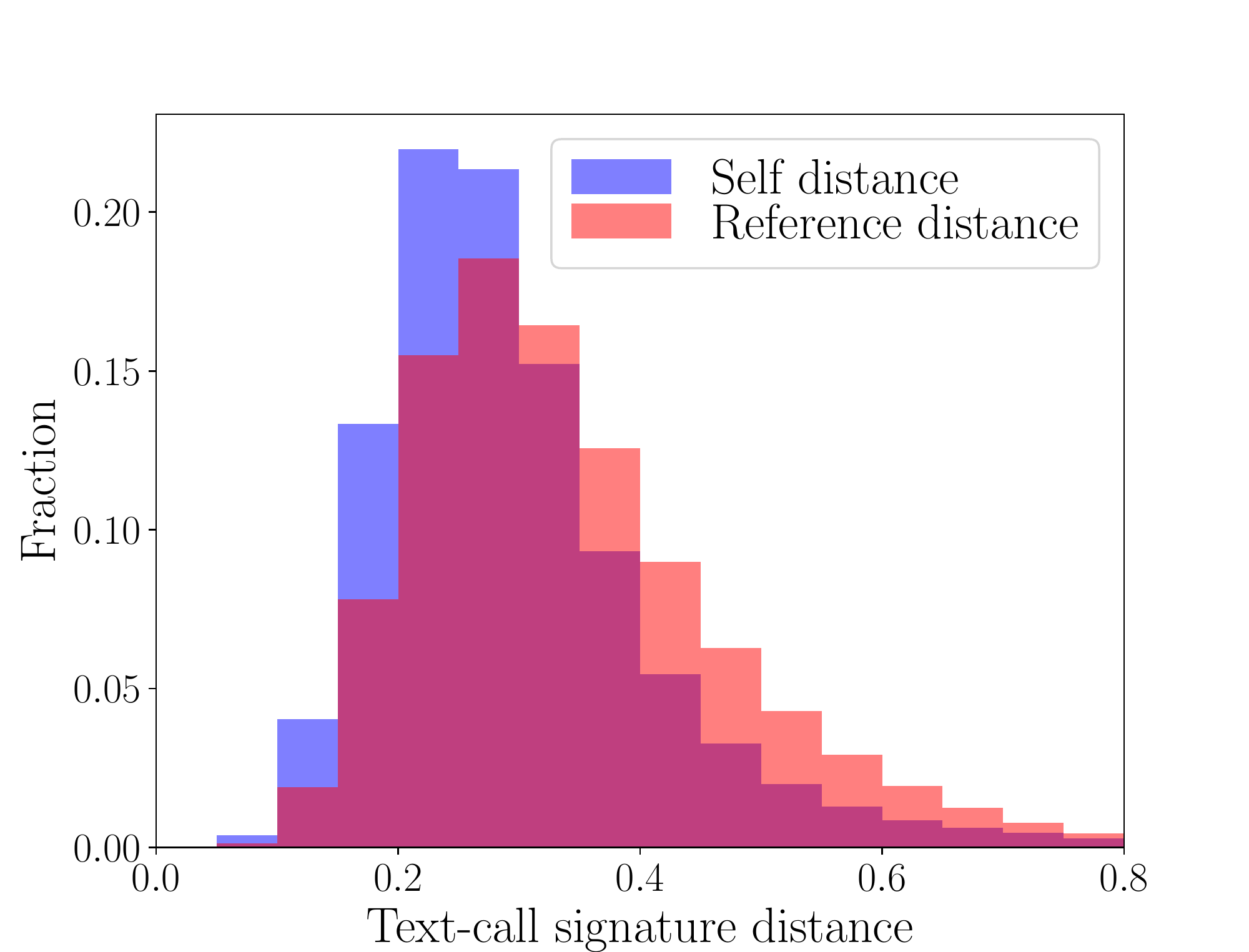}
  \caption{}
  \label{fig:large_t_c_similarity}
\end{subfigure}%
\begin{subfigure}{.45\textwidth}
  \centering
  \includegraphics[width=\linewidth]{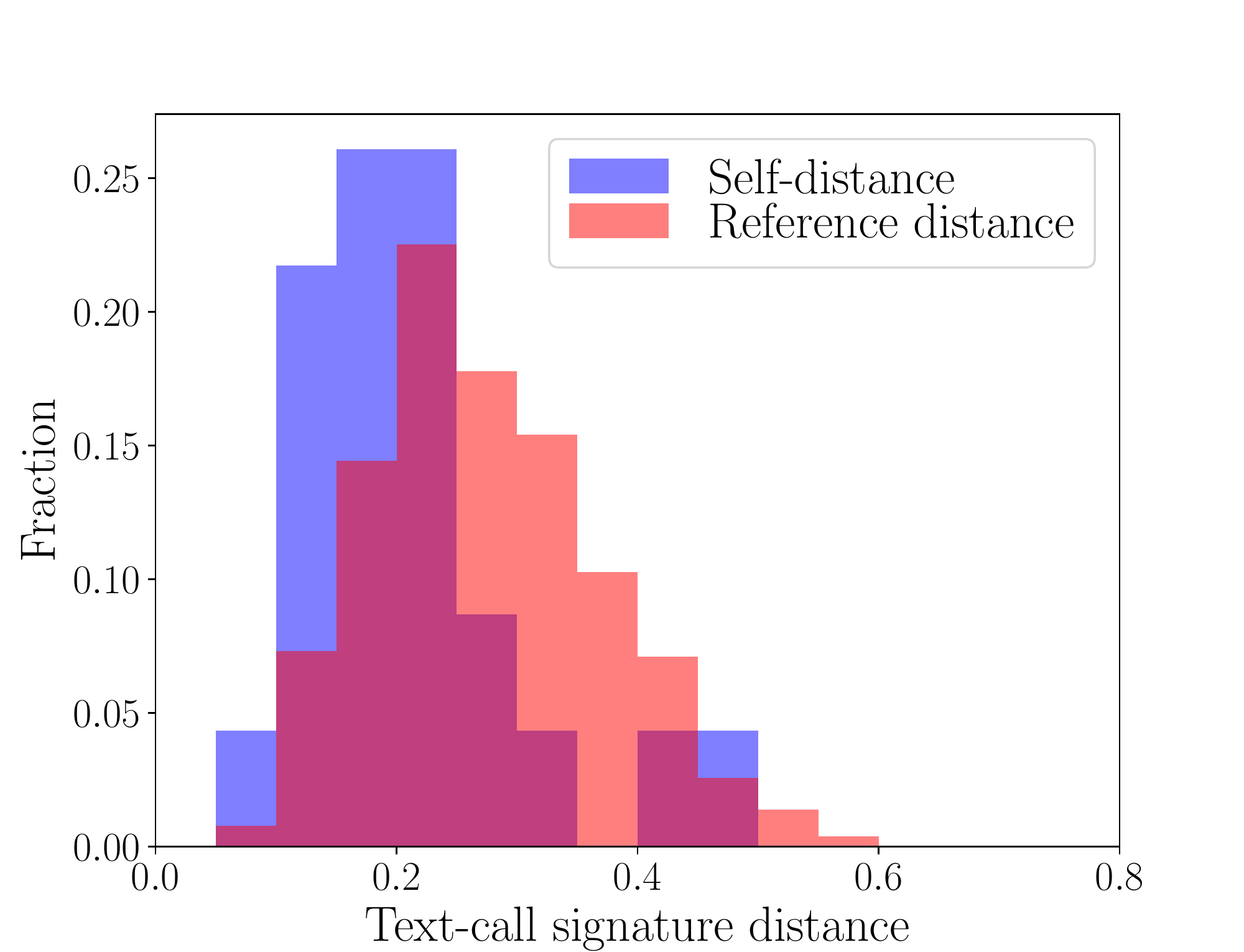}
  \caption{}
  \label{fig:oxford_t_c_similarity}
\end{subfigure}
\caption{\label{fig:all_t_c_similarity}The similarities of social signatures of different types. Panel (a) shows the call, text and mixed signatures of one person in the Dataset 1. The three signatures look similar. Panel (b) illustrates the average signatures over the population in Dataset 2. The population-level signatures are also fairly similar. Panels (c) and (d) compare the distance distributions of the call and text signatures of same egos with the distributions of call and text signatures of different people as a reference. The call and text signatures of each ego are more similar than pairs of signatures of different people.}
\end{figure*}

We have now established that the three types of signatures are persistent characteristics of egocentric networks. Next, we compare the shapes of these signatures, first at the population level and then at the level of individuals. It was shown in Ref.~\cite{saramaki2014persistence} that call signatures in DS2 are rather skewed: a small number of top-ranking alters get a disproportionate share of communications.  
We find that all three types of signatures show this skewed shape at the level of individuals and at the population level. This is seen in Fig.~\ref{fig:all_t_c_similarity} that shows the three types of signatures of one person (a) and the population-averaged signatures (b). 

It also appears that the two types of single-channel signatures are more similar for each ego than they are between egos--the shapes of the call and text signatures of an ego look similar. This is confirmed by use of the JSD. We calculated the self-distances between an ego's call and text signature as well as reference distances between all pairs of call and text signatures,
aggregated over the entire period of observation. The resulting distributions for both data sets again indicate that self-distances are on average smaller than reference distances (Fig.~\ref{fig:all_t_c_similarity}). Even though the difference is slightly less pronounced than for distances of the same signature type between different intervals (Fig.~\ref{fig:persistence_call_sig}), the shapes of call and text signatures of an ego appear to correlate.

\subsection{Single-Channel Egocentric Networks Differ in Composition}

The similarity in the shapes of the call and text signatures of each ego would be expected if their call and text networks were similar and included the same alters with similarly ranked weights. However, this is not the case: the call and text networks of an ego are typically different. Instead of the same alters appearing in both networks, many alters are only called or texted, and therefore included in one network only. This is in line with literature on network-level differences~\cite{nanavati2008,wang2013,zignani2014}.

\begin{figure*}
\centering
\begin{subfigure}{.45\textwidth}
  \centering
  \includegraphics[width=\linewidth]{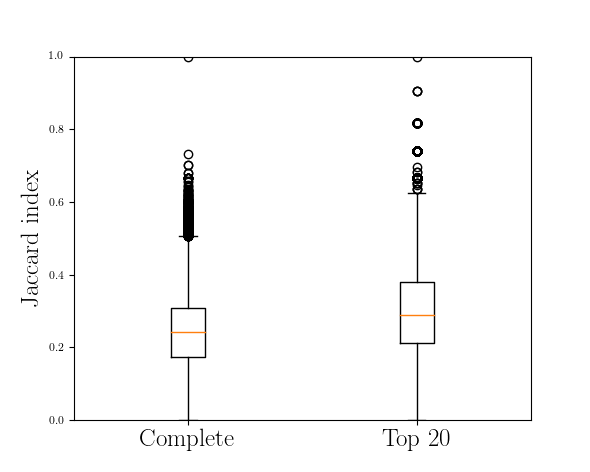}
  \caption{}
  \label{fig:big_text_call_jacc}
\end{subfigure}%
\begin{subfigure}{.45\textwidth}
  \centering
  \includegraphics[width=\linewidth]{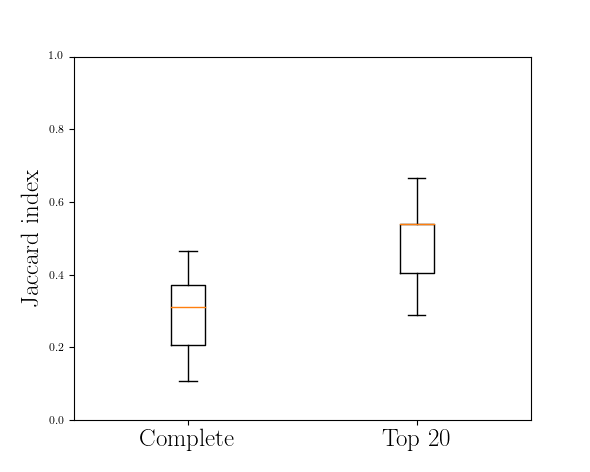}
  \caption{}
  \label{fig:oxford_text_call_jacc}
\end{subfigure}
\begin{subfigure}{.45\textwidth}
  \centering
  \includegraphics[width=\linewidth]{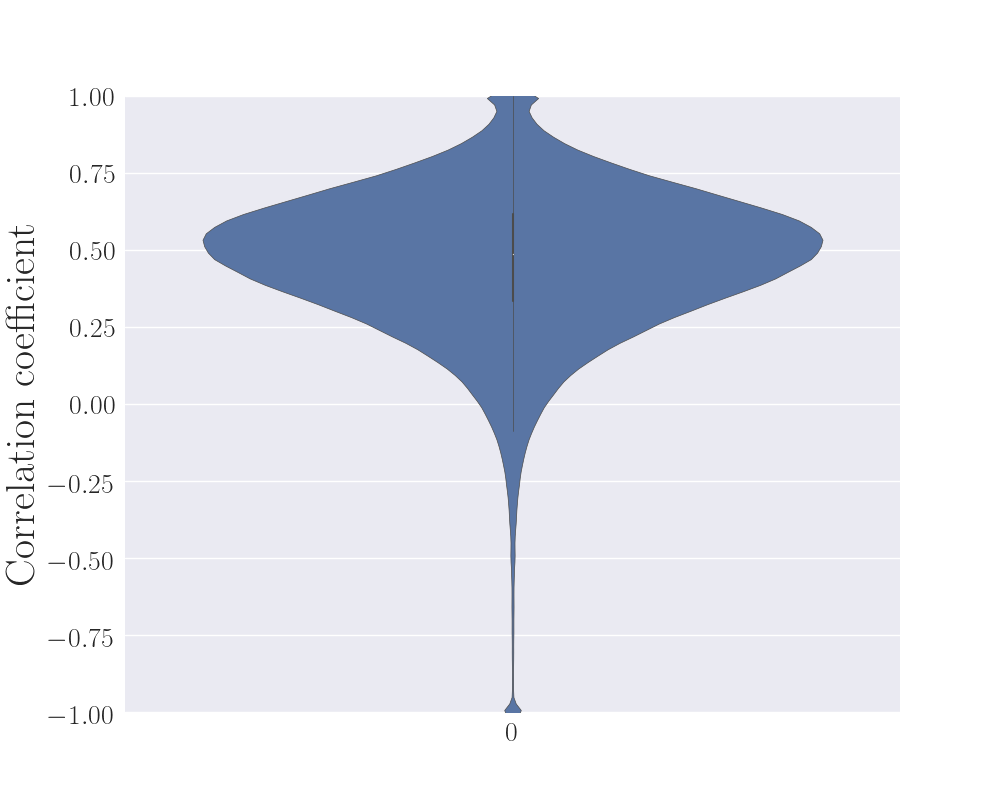}
  \caption{}
  \label{fig:big_text_call_corr}
\end{subfigure}%
\begin{subfigure}{.45\textwidth}
  \centering
  \includegraphics[width=\linewidth]{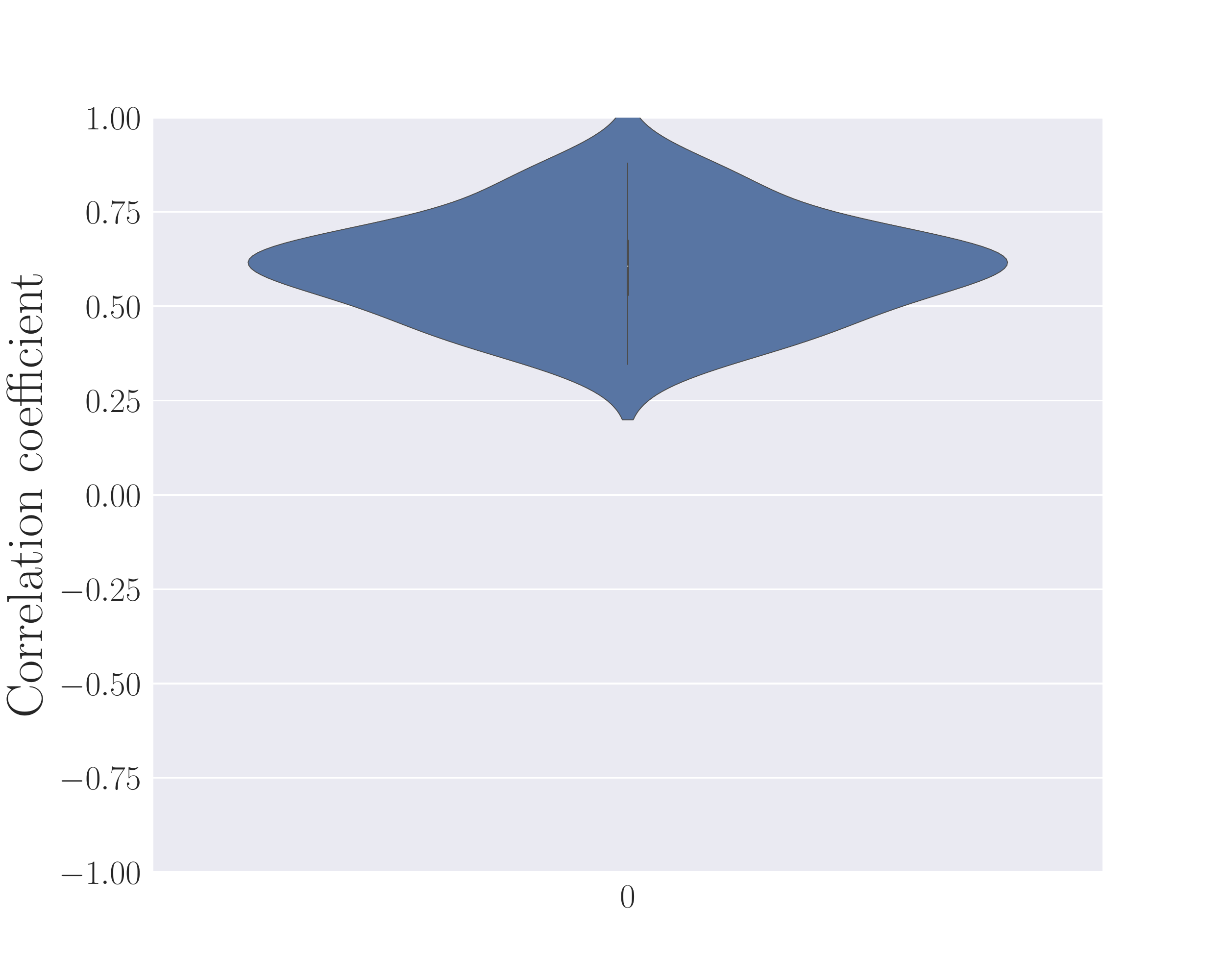}
  \caption{}
  \label{fig:oxford_text_call_corr}
\end{subfigure}
\caption{\label{fig:call_text_corr_and_jacc} Although the shapes of call and text signatures of an ego are relatively similar to each other, the egocentric networks formed through different channels are different in the membership and ranking of alters. The distribution of Jaccard indices between the sets of call and text alters are shown in (a) for DS1 and (b) for DS2. The distribution of correlation coefficients between call ranks and text ranks of those alters who are in both networks is shown in (c) for DS1 and in (d) for DS2. Alters who are only in one of the networks are not considered.}
\end{figure*}

This can be seen for both datasets in the distributions of Jaccard indices in Fig.~\ref{fig:call_text_corr_and_jacc} a) and b), computed for the sets of called and texted alters of each ego. The values of the Jaccard indices are mostly low. This means that while some alters are in both networks, most alters are not. Also, as seen in the lower panels (c,d) of Fig.~\ref{fig:call_text_corr_and_jacc}, the ranks of those alters who are present in both call and text networks correlate only moderately: an alter who is among the most called alters may receive a far smaller share of text messages.  

\subsection{Channel Choice Does not Depend on Alter Rank}

Next, we investigated whether there are systematic differences in the call and text networks of egos; such differences might explain the signature shapes and their similarities, despite call and text networks being different. To this end, we take a look at mixed egocentric networks calculated using both calls and texts, and investigate their weight composition. We focus on the share of calls and texts for each rank; note that since we are counting time slots, there are slots with both channels present.

\begin{figure*}
\includegraphics[width=0.9\linewidth]{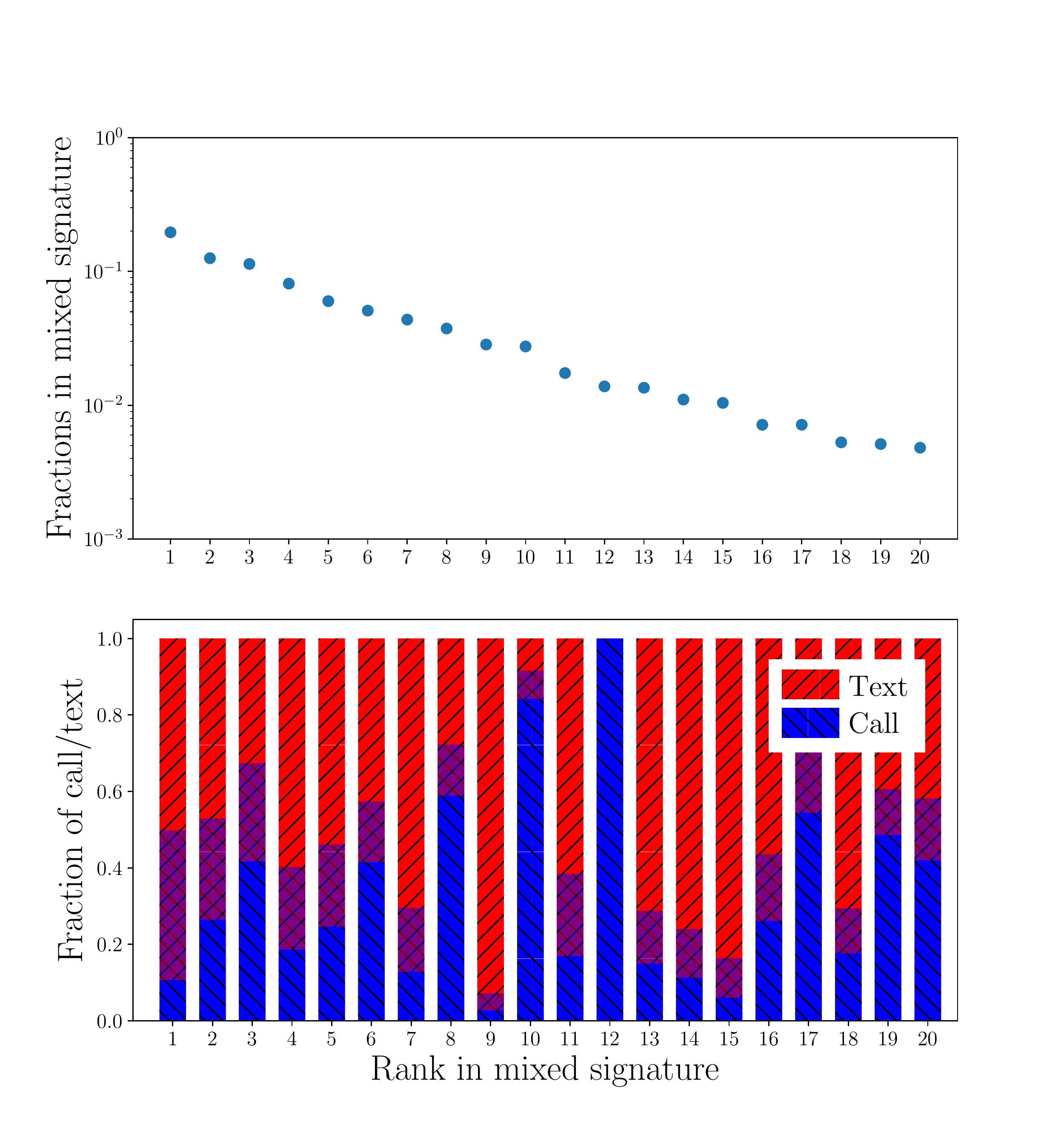}
\caption{Top panel: an example mixed signature for one of the egos in DS2. Bottom panel: fractions of calls and texts for the same ego, for each rank. The red and blue areas of the bars represent the fractions of texts and calls, respectively, of the link weight between the ego and the alter. The purple areas represent the fraction of time slots with both calls and texts.     
}
\label{fig:mixed_percentage_bars}
\end{figure*}

\begin{figure*}
\centering
\begin{subfigure}{.9\textwidth}
  \centering
  \includegraphics[width=0.9\linewidth]{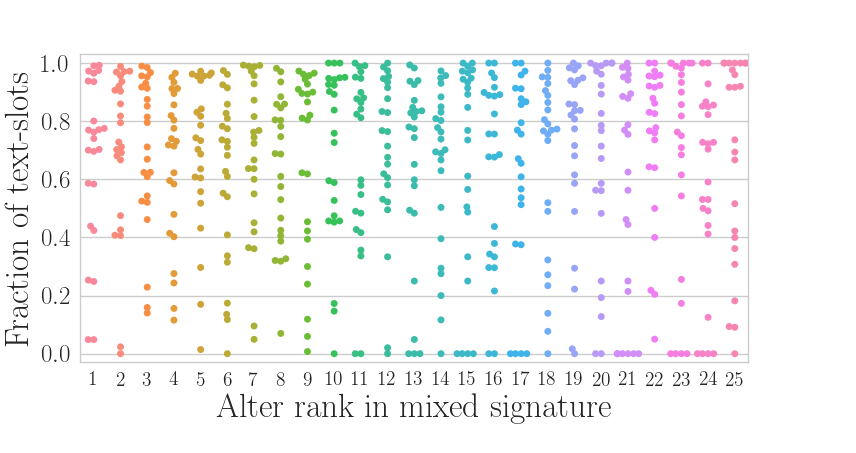}
  \caption{}
  \label{fig:swarmp_oxford}
\end{subfigure}
\begin{subfigure}{.9\textwidth}
  \centering
  \includegraphics[width=\linewidth]{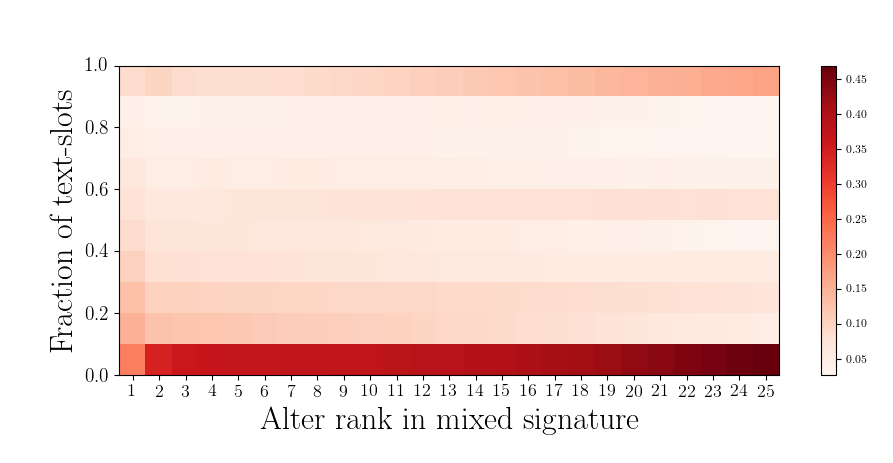}
  \caption{}
  \label{fig:big_call_fractions_heatmap}
\end{subfigure}
\caption{Top (a): Each dot shows the fraction of text slots in an ego-alter relationship as a function of rank, for the smaller DS2. The plot contains all ego-alter pairs. There appears to be no general pattern, except that the top ranks are mostly occupied by alters who are both called and texted, while in the tails of the signature (ranks $>$10 or so), there are more alters who are only texted or called. Bottom (b): A heat-map version of the top panel for DS1, with colors indicating the number of ego-alter pairs with a given fraction of texts at each rank in DS1. In this dataset, texts are used much less than in DS2.
}\label{fig:textshares}
\end{figure*}

One example mixed signature and its weight composition are shown in Fig.~\ref{fig:mixed_percentage_bars}. It appears that there is no clear pattern, except perhaps an slightly increased focus on calls around ranks 11-16. A likely explanation is that the choice of communication channel depends on features of the relationship in question other than its emotional closeness that correlates with ranks. This is supported by Fig.~\ref{fig:textshares} that shows the shares of text messages in all ego-alter relationships of DS2 (top) and a large sample of ego-alter relationships of DS1 (bottom). In DS2, the only systematic feature is that alters at top ranks typically receive both calls and text messages, and the fraction of text-only and call-only relationships increases towards the lower ranks. Top relationships appear more balanced regarding communication channels in DS1 too. Beyond that, there are no systematic changes that depend on alter rank.

\section{Discussion}

Social signatures quantify how people allocate their communication across the members of their personal networks. In this paper, we used two separate datasets to explore how social signatures based on calls, texts or both vary over time and across individuals. There were three key findings. First, individuals vary in how they allocate their time across their ego networks and this variation is persistent over time, despite a turnover of individuals alters in the network. This finding was initially reported in Ref.~\cite{saramaki2014persistence} using a small sample of 24 students of a similar age going through the specific transition from school to University or work (DS2 in this paper); Ref.~\cite{centegheller2017} confirmed the finding for a data set of $N=93$. The current paper replicates this finding in a much larger, more demographically diverse sample of over 500,000 people. Second, this individual variation in social signatures was present across different channels of communication - phone calls, text messages and a combined network based on both calls and texts. This is despite the fact that there was little overlap in the individuals who were called and texted. Third, regardless of the channel, the top alters get a disproportionately large fraction of all communication. This is seen in the shapes of three types of signatures (calls, texts, and mixed). Thus individual variation in social signatures does not appear to depend on the channel of communication or the specific alters in the network at a particular point in time, but instead reflects a stable characteristic of how individuals distribute their communication effort across their personal networks.

Why are the social signatures from different channels similar? One could envision an underlying complete egocentric network with tie strengths that measure the closeness of all the relationship an ego maintain with their alters. Within this network there is a distribution of tie strengths where there are few strong and many weak ties~\cite{dunbar2018anatomy}. Then observations on one channel of communication would be incomplete samples of the underlying complete network (see, e.g., \cite{nanavati2008,wang2013,zignani2014} for studies on network-level differences between calls and texts). Individuals differ in how they allocate their communication across their network, some individuals allocating a greater fraction of communication to a smaller number of alters and others allocating communication more evenly across their network. Thus constructing the social signatures based on different channels of communication would still pick up this individual variation, even if the specific alters detected by the different channels of communication vary. There may also be less fundamental reasons: communication habits, memory effects, or similar. However, it has been shown that for calls, alter ranks do correlate with emotional closeness~\cite{saramaki2014persistence}, which supports the first explanation. To fully understand which sample of alters is captured by different channels of communication, further research is needed on how people use different channels of communication to maintain their set of relationships to family and friends, and how this communication is related to the underlying tie strength of the relationship~\cite{reid2005,Vlahovic2012,wiese2015you}.

The finding that individuals have social signatures that are stable over time and persist despite the turnover of individual alters has now been shown in a number of samples from different countries and across different channels of communication including phone calls~\cite{saramaki2014persistence,centegheller2017}, text messages and combined call and text networks (this study) and email ~\cite{godoy2016long}. Given the robustness of this finding, further research is now needed on the causes and consequences of individual variation in social signatures. Whilst everyone is subject to similar fundamental time and cognitive constraints on sociality ~\cite{dunbar2018anatomy}, the way people choose to allocate their communication effort across their networks shows stable individual variation. Some of this individual variation appears to be due to personality characteristics~\cite{centegheller2017}, which are also broadly stable over time ~\cite{caspi2005personality}. Other characteristics that may be associated with individual variation in social signatures are age and gender, which are linked with variation in communication patterns ~\cite{bhattacharya2016sex} and friendship styles ~\cite{roberts2015managing}. Further, given the importance of social relationships to health and well-being \cite{house1988,lyubomirsky2005,wittig2008,holtlunstad2010,manninen2017} individual variation in social signatures may have consequences for outcomes such as stress and loneliness. Whilst all people distribute their communication very unevenly across their network~\cite{saramaki2014persistence}, some people focus an even greater proportion of their communication on a smaller number of alters. Further research could examine how these different patterns of time allocation across the network are linked to well-being, particularly during times of network change which put pressure on the time required to maintain relationships, such as the transition to University \cite{Roberts11} or entering into romantic relationships \cite{milardo1983developing}. Further, it would be important to see that our results can be replicated with other data sets containing calls and texts; as usual for this kind of data, our data cannot be made public because of privacy reasons. To conclude, this study demonstrated using two separate samples that there is individual variation in the way people allocate their time across their social networks, and these social signatures are persistent over time and across different channels of communication.

\section{Declarations}

\textbf{Availability of data and material.} The source data used in this study cannot be made publicly available because of privacy restrictions. Ref.~\cite{saramaki2014persistence} has some Supplementary Data in relation to DS2. 

\textbf{Competing interests.} The authors declare that they have no competing interests.

\textbf{Funding.} SH and JS acknowledge funding from the Academy of Finland, project n:o 297195.

\textbf{Acknowledgements.} We thank A.-L. Barab\'asi for providing data set DS1.

\textbf{Author contributions.} SGBR and RIMD conceived and performed the experiments; SH and JS analyzed data; SH, JS, SGBR, and RIMD wrote the paper.

\bibliography{citations}

\begin{thebibliography}{10}

\bibitem{house1988}
J.~House, K.~Landis, and D.~Umberson, ``Social relationships and health,'' {\em
  Science}, vol.~241, pp.~540--545, 1988.

\bibitem{lyubomirsky2005}
S.~Lyubomirsky, L.~King, and E.~Diener, ``The benefits of frequent positive
  affect: Does happiness lead to success?,'' {\em Psychological Bulletin},
  vol.~131, pp.~803--855, 2005.

\bibitem{wittig2008}
R.~Wittig, C.~Crockford, J.~Lehmann, P.~Whitten, R.~Seyfarth, and D.~Cheney,
  ``Focused grooming networks and stress alleviation in wild female baboons,''
  {\em Hormones and Behavior}, vol.~54, pp.~170--177.

\bibitem{holtlunstad2010}
J.~Holt-Lunstad, T.~Smith, and J.~Layton, ``Social relationships and mortality
  risk: A meta-analytic review,'' {\em PLoS Medicine}, vol.~7, p.~e1000316,
  2010.

\bibitem{manninen2017}
S.~Manninen, L.~Tuominen, R.~Dunbar, T.~Karjalainen, J.~Hirvonen, E.~Arponen,
  R.~Hari, I.~J\"a\"askel\"ainen, M.~Sams, and L.~Nummenmaa, ``Social laughter
  triggers endogenous opioid release in humans,'' {\em The Journal of
  Neuroscience}, vol.~37, pp.~6125--6131, 2017.

\bibitem{Granovetter}
M.~S. Granovetter, ``The strength of weak ties,'' {\em American Journal of
  Sociology}, vol.~79, pp.~1360--1380, 1973.

\bibitem{Burt}
R.~S. Burt, {\em Structural Holes: The Social Structure of Competition}.
\newblock Harvard University Press, 1995.

\bibitem{miritello2013time}
G.~Miritello, E.~Moro, R.~Lara, R.~Mart{\'\i}nez-L{\'o}pez, J.~Belchamber,
  S.~G. Roberts, and R.~I. Dunbar, ``Time as a limited resource: Communication
  strategy in mobile phone networks,'' {\em Social Networks}, vol.~35, no.~1,
  pp.~89--95, 2013.

\bibitem{miritello2013limited}
G.~Miritello, R.~Lara, M.~Cebrian, and E.~Moro, ``Limited communication
  capacity unveils strategies for human interaction,'' {\em Scientific
  reports}, vol.~3, 2013.

\bibitem{Roberts2009}
S.~G.~B. Roberts, R.~I.~M. Dunbar, T.~V. Pollet, and T.~Kuppens, ``Exploring
  variation in active network size: Constraints and ego characteristics,'' {\em
  Social Networks}, vol.~31, no.~2, pp.~138--146, 2009.

\bibitem{Onnela2007}
J.~P. Onnela, J.~Saram\"{a}ki, J.~Hyv\"{o}nen, G.~Szab\'{o}, D.~Lazer,
  K.~Kaski, J.~Kert\'{e}sz, and A.~L. Barab\'{a}si, ``Structure and tie
  strengths in mobile communication networks,'' {\em Proceedings of the
  National Academy of Sciences}, vol.~104, no.~18, pp.~7332--7336, 2007.

\bibitem{saramaki2014persistence}
J.~Saram{\"a}ki, E.~A. Leicht, E.~L{\'o}pez, S.~G. Roberts, F.~Reed-Tsochas,
  and R.~I. Dunbar, ``Persistence of social signatures in human
  communication,'' {\em Proceedings of the National Academy of Sciences},
  vol.~111, no.~3, pp.~942--947, 2014.

\bibitem{centegheller2017}
S.~Centellegher, E.~L\'opez, J.~Saram\"aki, and B.~Lepri, ``Personality traits
  and ego-network dynamics,'' {\em PLOS ONE}, vol.~12, pp.~1--17, 03 2017.

\bibitem{Vlahovic2012}
T.~A. Vlahovic, S.~Roberts, and R.~Dunbar, ``Effects of duration and laughter
  on subjective happiness within different modes of communication,'' {\em
  Journal of Computer-Mediated Communication}, vol.~17, pp.~436--450, 2012.

\bibitem{reid2005}
D.~J. Reid and F.~Reid, ``Textmates and text circles: Insights into the social
  ecology of {SMS} text messaging,'' in {\em Mobile World. Computer Supported
  Cooperative Work} (L.~Hamill, L.~A, and D.~D, eds.), London: Springer, 2005.

\bibitem{nanavati2008}
A.~A. Nanavati, R.~Singh, D.~Chakraborty, K.~Dasgupta, S.~Mukherjea, G.~Das,
  S.~Gurumurthy, and A.~Joshi, ``Analyzing the structure and evolution of
  massive telecom graphs,'' {\em IEEE Transactions on Knowledge and Data
  Engineering}, vol.~20, no.~5, pp.~703--718, 2008.

\bibitem{wang2013}
Y.~Wang, M.~Faloutsos, and H.~Zang, ``On the usage patterns of multimodal
  communication: Countries and evolution,'' in {\em 2013 Proceedings IEEE
  INFOCOM}, pp.~3135--3140, 2013.

\bibitem{zignani2014}
M.~Zignani, C.~Quadri, S.~Bernardinello, S.~Gaito, and G.~P. Rossi, ``Calling
  and texting: Social interactions in a multidimensional telecom graph,'' in
  {\em 2014 Tenth International Conference on Signal-Image Technology and
  Internet-Based Systems}, pp.~408--415, 2014.

\bibitem{quadri2014}
C.~Quadri, M.~Zignani, L.~Capra, S.~Gaito, and G.~P. Rossi, ``Multidimensional
  human dynamics in mobile phone communications,'' {\em PLOS ONE}, vol.~9,
  no.~7, pp.~1--12, 2014.

\bibitem{saramaki2015seconds}
J.~Saram\"aki and E.~Moro, ``{From seconds to months: an overview of
  multi-scale dynamics of mobile telephone calls},'' {\em The European Physical
  Journal B: Condensed Matter and Complex Systems}, vol.~88, pp.~1--10, 2015.

\bibitem{Karsai2011}
M.~Karsai, M.~Kivel\"a, R.~K. Pan, K.~Kaski, J.~Kert\'esz, A.-L. Barab\'asi,
  and J.~Saram\"aki, ``Small but slow world: How network topology and
  burstiness slow down spreading,'' {\em Phys. Rev. E}, vol.~83, p.~025102,
  2011.

\bibitem{kivela2012multiscale}
M.~Kivel{\"a}, R.~K. Pan, K.~Kaski, J.~Kert{\'e}sz, J.~Saram{\"a}ki, and
  M.~Karsai, ``Multiscale analysis of spreading in a large communication
  network,'' {\em J. Stat. Mech. Theor. Exp.}, vol.~2012, no.~03, p.~P03005,
  2012.

\bibitem{Roberts11}
S.~G.~B. Roberts and R.~I.~M. Dunbar, ``The costs of family and friends: an
  18-month longitudinal study of relationship maintenance and decay,'' {\em
  Evolution and Human Behavior}, vol.~32, pp.~186--197, 2011.

\bibitem{backlund2014}
V.-P. Backlund, J.~Saram\"aki, and R.~K. Pan, ``Effects of temporal
  correlations on cascades: Threshold models on temporal networks,'' {\em Phys.
  Rev. E}, vol.~89, p.~062815, 2014.

\bibitem{lin1991divergence}
J.~Lin, ``Divergence measures based on the shannon entropy,'' {\em IEEE
  Transactions on Information theory}, vol.~37, no.~1, pp.~145--151, 1991.

\bibitem{dunbar2018anatomy}
R.~Dunbar, ``The anatomy of friendship,'' {\em Trends in cognitive sciences},
  vol.~22, no.~1, pp.~32--51, 2018.

\bibitem{wiese2015you}
J.~Wiese, J.-K. Min, J.~I. Hong, and J.~Zimmerman, ``You never call, you never
  write: Call and {SMS} logs do not always indicate tie strength,'' in {\em
  Proceedings of the 18th ACM conference on computer supported cooperative work
  \& social computing}, pp.~765--774, ACM, 2015.

\bibitem{godoy2016long}
A.~Godoy-Lorite, R.~Guimer{\`a}, and M.~Sales-Pardo, ``Long-term evolution of
  email networks: statistical regularities, predictability and stability of
  social behaviors,'' {\em PloS one}, vol.~11, no.~1, p.~e0146113, 2016.

\bibitem{caspi2005personality}
A.~Caspi, B.~W. Roberts, and R.~L. Shiner, ``Personality development: Stability
  and change,'' {\em Annu. Rev. Psychol.}, vol.~56, pp.~453--484, 2005.

\bibitem{bhattacharya2016sex}
K.~Bhattacharya, A.~Ghosh, D.~Monsivais, R.~I. Dunbar, and K.~Kaski, ``Sex
  differences in social focus across the life cycle in humans,'' {\em Royal
  Society open science}, vol.~3, no.~4, p.~160097, 2016.

\bibitem{roberts2015managing}
S.~B. Roberts and R.~I. Dunbar, ``Managing relationship decay,'' {\em Human
  Nature}, vol.~26, no.~4, pp.~426--450, 2015.

\bibitem{milardo1983developing}
R.~M. Milardo, M.~P. Johnson, and T.~L. Huston, ``Developing close
  relationships: Changing patterns of interaction between pair members and
  social networks.,'' {\em Journal of Personality and Social Psychology},
  vol.~44, no.~5, p.~964, 1983.

\end{thebibliography}

\end{document}